\documentstyle[12pt]{article}
\begin{document}
\begin{center}
\vfill
\large\bf{Summation of Higher Order Effects\\
using the Renormalization Group Equation}
\end{center}
\vfill
\begin{center}
V. Elias, 
D.G.C. McKeon\\
Department of Applied Mathematics\\
University of Western Ontario\\
London, Ontario, Canada\\
N6A 1B7\\
\vspace{2cm}
T.N. Sherry\\
Department of Mathematical Physics\\
National University of Ireland Galway\\
Galway, IRELAND
\end{center}
\vspace{.2cm}
email:  velias@uwo.ca\\
dgmckeo2@uwo.ca\\
tom.sherry@nuigalway.ie
\begin{flushright}
\end{flushright}
tel. 519-661-2111, x88789\\
fax  519-661-3523
\eject
\section{Abstract}

The renormalization group (RG) is known to provide information about radiative corrections beyond the order in perturbation theory to which one has calculated explicitly. We first demonstrate the effect of the renormalization scheme used on these higher order effects determined by the RG. Particular attention is payed to the relationship between bare and renormalized quantities. Application of the method of characteristics to the RG equation to determine higher order effects is discussed, and is used to examine the free energy in thermal field theory, the relationship between the bare and renormalized coupling and the effective potential in massless scalar electrodynamics.

\section{Introduction}

The renormalization group (RG) exploits the fact that physical processes cannot have dependence on the mass paramter $\mu$ that enters in the course of computing radiative effects. This has two consequences. First of all, since explict dependence of a physical process on $\mu$ must be compensated by an implicit dependence through dependence of couplings, masses and wave functions on $\mu$, these quantities now ``run'' (viz. they depend on $\mu$ in a prescribed way).

A second consequence that seems to have been exploited only recently [1,2,3] is that the RG serves to determine portions of perturbative results beyond the order to which exploit calculations have been done; one-loop calculations fix ``leading log'' (LL) effects to all orders of perturbation theory, a two-loop calculation fixes ``next to leading log'' (NLL) effects etc.

In this paper we consider several aspects of this summation procedure using the RG equation. First, the renormalization scheme dependence of these summed quantities is analyzed. The relationship between bare and renormalized quantities (couplings, masses and fields) is discussed in an arbitrary scheme. The formal solution of the RG equation is examined and use of the method of characteristics to solve the RG equation is detailed. This method of characteristics is applied to computing the free energy in thermal field theory, the relationship between the bare and renormalized couplings and to the effective potential.

\section{Renormalization Scheme Dependence}

Dimensional regularization [4-6], combined with minimal subtraction (MS) [7] has proven to be efficient when computing radiative effects. In this approach, bare quantities appearing in the $n$-dimensional Lagrangian are expanded in terms of poles in $\epsilon = 2 - n/2$. In general, a bare coupling $g_B$ and the corresponding renormalized coupling $g$ are related in a mass independent renormalization scheme by
$$g_B = \mu^\epsilon \sum_{\nu = 1}^\infty \,\frac{a_\nu(g)}{\epsilon^\nu}\eqno(1)$$
where $\mu$ is a renormalization induced scale parameter. With MS [7], the first term in the expansion of eq. (1), $a_0(g)$, is chosen so that
$$a_0(g) = g.\eqno(2)$$

In general, explicit calculations leads to the expansion
$$a_\nu (g) = \sum_{k=\nu}^\infty \,a_{k,\nu} g^{2k+1}\eqno(3)$$
with
$$a_{0,0} = 1.\eqno(4)$$

In ref. [8], the sum of eq. (1) is reorganized so that
$$g_B = \mu^\epsilon \sum_{k=0}^\infty \,g^{2k+1} S_k \left(\frac{g^2}{\epsilon}\right)\eqno(5)$$
with
$$S_k(\xi)  = \sum_{\ell = 0}^\infty a_{k+\ell , \ell} \xi^\ell \eqno(6)$$
and
$$S_k(0) = a_{k,0}\;.\eqno(7)$$

Since $g_B$ is independent of $\mu$,
$$\mu \frac{dg_B}{d\mu} = 0 = \left(\mu \frac{\partial}{\partial\mu} + \mu \frac{dg}{d\mu}\,\frac{\partial}{\partial g}\right)g_B\,.\eqno(8)$$
If the expansion of eq. (1) is substituted into eq. (8), terms $O(\epsilon)$ cancel provided
$$\mu \frac{dg}{d\mu} = -\epsilon \frac{a_0(g)}{a_0^\prime (g)} + \beta(g)\eqno(9)$$
where $\beta(g)$ is the usual $\beta$ function that arises in the $\epsilon \rightarrow 0$ limit. Terms of order $\epsilon^0$ cancel if
$$a_1 - \frac{a_0}{a_0^\prime} a_1^\prime + \beta a_0^\prime = 0\eqno(10)$$
so that
$$\beta (g) = \frac{a_0^2}{a_0^{\prime 2}} \,\frac{d}{dg} \left(\frac{a_1}{a_0}\right),\eqno(11)$$
showing that $a_0(g)$ (which is dictated by the renormalization scheme) and $a_1(g)$ (which must be computed) determine $\beta (g)$. Terms of order $\epsilon^{-n} (n > 0)$ in eq. (1) serve to fix $a_2$, $a_3$ etc. in terms of $a_0$ and $a_1$.

If now we define
$$\tilde{g} = a_0(g)\eqno(12)$$
then $\tilde{g}$ is the coupling in MS. It is evident that by eq. (9) and the chain rule
$$\mu \frac{d\tilde{g}}{d\mu} = -\epsilon \tilde{g} + \tilde{\beta}(\tilde{g}) = \frac{d\tilde{g}}{dg} \left( \mu \frac{dg}{d\mu}\right)\eqno(13)$$
$$= a_0^\prime (g) \left(-\epsilon \frac{a_0(g)}{a_0^\prime (g)} + \beta (g)\right)\nonumber$$
so that
$$\tilde{\beta} (a_0(g)) = a_0^\prime (g) \beta (g).\eqno(14)$$

We can now generalize the results of ref. (8) to an arbitrary renormalization scheme. By substituting eq. (5) into eq. (1) we see that if $U = g^2/\epsilon$, then
$$\sum_{n=0}^\infty g^{2n} \left[ \frac{g^3}{U} S_n(U) + \left(-\frac{a_0}{a_0^\prime} \,\frac{g^2}{U} + \beta(g)\right)\left((2n+1)S_n (U) + 2U S_n^\prime (U)\right)\right] = 0.\eqno(15)$$
Using eq. (3), we see that
$$\frac{a_0(g)}{a_0^\prime (g)} = \frac{g + a_{10}g^3 + a_{20}g^5 + \ldots}{1 + 3a_{10}g^2 + 5a_{20}g^4 + \ldots}
\eqno(16)$$
$$\equiv g + \alpha_3g^3 + \alpha_5g^5 +\ldots\eqno(17)$$
Furthermore, if we make the expansion
$$\beta(g) = \sum_{k=1}^\infty B_{2k+1} g^{2k+1}\eqno(18)$$
then together eqs. (17) and (18) reduce eq. (15) to a set of nested ordinary differential equations. In particular, $S_0$ and $S_1$ satisfy
$$\left(1 - UB_3\right) S_0^\prime - \frac{1}{2} B_3 S_0 = 0\eqno(19)$$
$$\frac{1}{U} S_1 + \left( - \frac{1}{U} + B_3\right) \left(3 S_1 + 2US_1^\prime\right)
+ \left(-\frac{\alpha_3}{U} + B_5\right)\left(S_0 + 2U S_0^\prime\right) = 0\eqno(20)$$
whose solutions; subject to the boundary condition of eq. (7), are
$$S_0(U) = W^{-1/2}\eqno(21)$$
and
$$S_1(U) = \frac{1}{W^{1/2}(1-W)} \left[ \left(\frac{\alpha_3}{2} - \frac{B_5}{2B_3}\right)\left(-1 + W^{-1}\right) - \frac{B_5}{2B_3} \ln W\right]\eqno(22)$$
where $W = 1 - B_3 U$. It is apparent from eqs. (21) and (22) that as $\epsilon \rightarrow 0$ (ie, $U \rightarrow \infty$ $W \rightarrow \infty$ ),
$$S_0 \rightarrow 0,\;\;\;\; S_1 \rightarrow 0\;.\eqno(23)$$
This is consistent with the result
$$\lim_{\epsilon \rightarrow 0} S_k (U) = 0\eqno(24)$$
found when using the MS scheme [8]. This can also be shown by deriving a closed form expression for $g_B$ in terms of $\epsilon$ and $g$.

To do this, we note that at $O(\epsilon^{-k})$, eq. (8) gives
$$-\frac{a_0^2}{a_0^\prime}\,\frac{d}{dg} \left(\frac{a_k}{a_0}\right) + \beta a_{k-1}^\prime = 0\eqno(25)$$
so that
$$a_{k+1}(g) = a_0(g) \int_0^g d\lambda \frac{\beta(\lambda)a_k^\prime (\lambda)a_0^\prime(\lambda)}{a_0^2(\lambda)}\,, \eqno(26)$$
thus fixing $a_{k+1}$ iteratively. Furthermore, from eq. (25)
$$-\frac{a_0}{a_0^\prime} a_k^\prime + a_k + \beta a_{k-1}^\prime = 0,\eqno(27)$$
which upon multiplying by $\epsilon^{-k+1}$ and summing over $k$ gives
$$\epsilon\left(g_B - \frac{a_0(g)}{a_0^\prime(g)} \,\frac{\partial}{\partial g} g_B\right) + \beta (g) \frac{\partial g_B}{\partial g} = 0,\eqno(28)$$
as expected from eqs. (1), (8) and (9). A formal solution of eq. (28) is
$$g_B = \exp \left( - \int^g dx \frac{\epsilon}{\left(\beta (x) - \epsilon \frac{a_0(x)}{a_0^\prime(x)}\right)} + K\right).\eqno(29)$$
The constant $K$ is chosen so that eq. (8) is satisfied, leading to
$$g_B = \mu^\epsilon \exp \left( - \int_\kappa^g dx \frac{\epsilon}{\left(\beta (x) - \epsilon \frac{a_0(x)}{a_0^\prime(x)}\right)} \right) \eqno(30)$$
where $\kappa$ is a cut off to render the integral in eq. (30) convergent. Consistency with $S_0$ and $S_1$ in eqs. (21) and (22) is obtained by having
$$g_B = \mu^\epsilon g \exp \left( - \int_0^g dx \left[\frac{\epsilon}{\left(\beta (x) - \epsilon \frac{a_0(x)}{a_0^\prime(x)}\right)} + \frac{1}{x}\right]\right).\eqno(31)$$
(This eq. is similar to eq. (7.5) of ref. [7].) From eq. (31), if we consider the limit $\epsilon \rightarrow 0$, then 
$$\lim_{\epsilon \rightarrow 0} g_B = g \exp \left(- \lim_{\delta \rightarrow 0^+} \int_\delta^g \frac{dx}{x}\right) = 0,\eqno(32)$$
showing that $g_B$ vanishes as $\epsilon \rightarrow 0$ for any renormalization scheme, which is consistent with the MS result of ref. [8].

We now directly consider the dependence of $g$
on the renormalization scheme. In ref. [9] the renormalization scheme is characterized by the coefficients $B_7, B_9 \ldots$ of eq. (18). (From eq. (14) it is evident that $B_3$ and $B_5$ are unaltered by changes in $a_{k0}$ and hence are renormalization scheme independent.) We shall use the coefficients $a_{k0}$ themselves to characterize the renormalization scheme.

In ref. [9] where scheme dependence is characterized by $B_{2k+5} (k = 1,2 \ldots)$, one can define functions $\beta_{2k+5} (g, B_{2j +5})$ by
$$\frac{\partial g}{\partial B_{2k+5}} = \beta_{2k+5} \left(g, B_{2j+5}\right).\eqno(33)$$
When $\epsilon = 0$, eq. (9) reduces to
$$\mu \frac{\partial g}{\partial \mu} = \beta \left(g, B_{2k+1}\right).\eqno(34)$$
Now
$$
\mu \frac{\partial^2 g}{\partial \mu \partial B_{2k+5}} = \mu \frac{\partial^2 g}{\partial B_{2k+5} \partial\mu}\eqno(35)$$
and hence by eq. (18)
$$\frac{\partial \beta_{2k+5}}{\partial g} \beta = g^{2k+5} + \frac{\partial \beta}{\partial g} \beta_{2k+5},\eqno(36)$$
which can be integrated to give
$$\beta_{2k+5} \left(g, B_{2j+5}\right) = \beta \left( g, B_{2j+1}\right) \int_0^g dx\frac{x^{2k+5}}{\beta^2\left(x, B_{2j+1}\right)}.\eqno(37)$$
The explicit dependence of $\beta_{2k+5}$ on $B_{2j+5}$ makes integration of eq. (33) much more difficult than integration of eq. (34).

If we were to use $a_{k0}$ to characterize the renormalization scheme, then
$$\frac{dg_B}{da_{k0}} = 0.
\eqno(38)$$
If now eqs. (1) and (3) are substituted into eq. (38), at $O(\epsilon^0)$ we find that
$$\frac{\partial g}{\partial a_{j,0}} \equiv A_j \left(g, a_{j,0}\right) = \frac{-g^{2j+1}}{\left(\frac{\partial a_0}{\partial g} \left( g, a_{k,0}\right)\right)}\,.\eqno(39)$$
This equation is not separable and hence cannot be integrated. Poles of order $\epsilon^{-\nu}(\nu \geq 1)$ in eq.  (38) show how $a_{k,\nu}(k \geq \nu \geq 1)$ depend on $a_{j,0}$ as we have
$$\sum_{k = \nu}^\infty \frac{\partial a_{k,\nu}}{\partial a_{j,0}} \,\frac{\partial a_\nu}{\partial a_{k,\nu}} + \frac{\partial g}{\partial a_{j,0}} \,\frac{\partial a_\nu}{\partial g} = 0. \eqno(40)$$
Eqs. (3), (39) and (40) together lead to
$$\sum_{k = \nu}^\infty \left(
\frac{\partial a_{k,\nu}}{\partial a_{j,0}} g^{2k+1} + A_j(2k + 1) a_{k,\nu} g^{2k}\right) = 0.
\eqno(41)$$
From eq. (39), it is apparent that $A_j$ is of order $g^{2j+1}$, and hence if
$$\frac{\partial a_{k,\nu}}{\partial a_{j,0}} = \sum_{\ell = j}^\infty \lambda_\ell^{(k,\nu)} g^{2\ell}\;,\eqno(42)$$
eq. (41) can be used to determine the coefficients $\lambda_\ell^{(k,\nu)} (\ell = j, j + 1, \ldots)$.

We now examine the relationship between the bare and renormalized masses and fields in the context of dimensional regularization and a mass independent renormalization scheme.

Taking the bare and renormalized mass to have the same mass dimension, then we have
$$m_B = Z_m(g)m = \sum_{\nu = 0}^\infty \frac{c_\nu(g)}{\epsilon^\nu} m.\eqno(43)$$
With the MS scheme, $c_0(g) = 1$; in general
$$c_\nu (g) = \sum_{k = \nu}^\infty c_{k,\nu} g^{2k}\;,\eqno(44)$$
with $c_0 = 1$. We now take
$$\mu \frac{dm_B}{d\mu} = 0 = \left( -\epsilon \frac{a_0}{a_0^\prime} + \beta\right) \frac{\partial m_B}{\partial g} + \left(\epsilon \frac{a_0}{a_0^\prime} \,\frac{c_0^\prime}{c_0} + \gamma_m\right)m\,\frac{\partial m_B}{\partial m}\,.\eqno(45)$$
Where
$$\mu \frac{\partial m}{\partial \mu} = + \epsilon \frac{a_0}{a_0^\prime} \,\frac{c_0^\prime}{c_0} + \gamma_m\eqno(46)$$
and $\mu \frac{\partial g}{\partial\mu}$ is given by eq. (9). Terms of order $\epsilon$ in eq. (45) cancel at order $\epsilon^0$ we find that $\gamma_m$ is fixed by the equation
$$- \frac{a_0}{a_1^\prime} c_1^\prime + \beta c_0^\prime +
\frac{a_0}{a_0^\prime} \,\frac{c_0^\prime}{c_0} c_1 + \gamma_m c_0 = 0.\eqno(47)$$
In the MS scheme, this reduces to $\gamma_m = g c_1^\prime$; in general $\gamma_m$ is determined by $a_0$, $a_1$, $c_0$ and $c_1$. At order $\epsilon^{-n}$, eq. (45) reduces to the consistency condition
$$- \frac{a_0}{a_0^\prime} c_{n+1}^\prime + \beta c_n^\prime + \frac{a_0}{a_0^\prime} \,\frac{c_0^\prime}{c_0} c_{n+1}  + \gamma_m c_n = 0.\eqno(48)$$

We now reorganize the sums in eqs. (43) and (44) in a manner analogous to eqs. (5) and (6). This involves defining
$$T_n (U) = \sum_{m=0}^\infty c_{n+m,m} U^m\eqno(49)$$
so that eq. (43) becomes
$$m_B = \sum_{n=0}^\infty g^{2n} T_n\left(\frac{g^2}{\epsilon}\right) m.\eqno(50)$$
Subsituting eq. (50) into eq. (45) leads to
$$\left[ -\frac{g^2}{U} \left( g + \alpha_3 g^3 + \alpha_5 g^5 + \ldots\right) + \left( B_3g^3 + B_5g^5 + \ldots\right)\right]\nonumber$$
$$\left[\;\sum_{k=0}^\infty 2g^{2k-1} \left(k T_k (U) + U T_k^\prime (U)\right)\right]\nonumber$$
$$+ \left[ \frac{g^2}{U} \left(\zeta_2 g^2 + \zeta_4 g^4 + \ldots\right) + \left( G_2 g^2 + G_4 g^4 + \ldots\right)\right]\nonumber$$
$$\left[\sum_{k=0}^\infty g^{2k} T_k(U)\right] = 0\eqno(51)$$
where $U = g^2/\epsilon$.

In eq. (51), the expansions of eqs. (17) and (18) as well as
$$\frac{a_0}{a_0^\prime}\,\frac{c_0^\prime}{c_0} = \sum_{k=1}^\infty \,\zeta_{2k} g^{2k}\eqno(52)$$
and
$$\gamma_m (g) = \sum_{k=1}^\infty G_{2k} g^{2k}\eqno(53)$$
have been used. At order $g^{2k+2}$, an equation relating $T_k$, $T_{k-1} \ldots T_0$ is found; these can be solved iteratively upon using the bondary condition
$$T_k(0) = c_{k,0}.\eqno(54)$$
In particular, at $0(g^2)$ eq. (51) leads to
$$\left(-\frac{1}{U} + B_3\right)\left(2UT_0^\prime\right) + G_2 T_0 = 0\eqno(55)$$
so that
$$T_0(U) = \left(1 - B_3 U\right)^{- G_2/2B_3}.\eqno(56)$$
The relative sign of $G_2$ and $B_3$ dictates the behaviour of $T_0(U)$ as $\epsilon \rightarrow 0$ (ie, $U \rightarrow \infty$). This is in contrast to the behaviour of the bare coupling as given in eq. (32).

The consistency condition of eq. (48) can be used to find a closed form expression for $m_B$ analogous to eq. (31).  Since by eq. (43) $m\frac{\partial m_B}{\partial m} = m_B$, eq. (45) can be written as
$$\left( - \frac{a_0}{a_0^\prime} \epsilon + \beta\right) \frac{dm_B}{dg} + \left( \frac{a_0}{a_0^\prime} \,\frac{c_0^\prime}{c_0} \epsilon + \gamma_m\right) m_B = 0.\eqno(57)$$
This equation is separable; solving it gives
$$m_B = m \exp - \int_0^g dx\left[\left(  \frac{a_0}{a_0^\prime} \,\frac{c_0^\prime}{c_0} \epsilon + \gamma\right)/ \left(- \frac{a_0}{a_0^\prime} \epsilon + \beta\right)\right].\eqno(58)$$
The integral over $x$ in eq. (58) is not singular provided $\epsilon \neq 0$; however if $\epsilon = 0$, a divergence appears that is consistent with the behaviour of $T_0$ in eq. (56) as $\epsilon \rightarrow 0$.

The relationship between $m_B$ and $m$ is similar to that between $A_B$ and $A$, the bare and renormalized fields respectively. Since the canonical dimension of a Bosonic field $A_B$ is $\mu^{1-\epsilon}$, we have in a mass independent renormalization scheme
$$A_B = \mu^{-\epsilon} \sum_{\nu = 0}^\infty \frac{d_\nu(g)}{\epsilon^\nu} A.\eqno(59)$$
We now take
$$\mu\frac{dA}{d\mu} = \left[ \left( 1 + \frac{a_0}{a_0^\prime}\,\frac{d_0^\prime}{d_0}\right)\epsilon + \gamma\right]A.\eqno(60)$$
Together, eqs. (9) and (60) show that eq. (59) leads to the RG equation
$$\mu \frac{dA_B}{d\mu} = 0 = \left[\mu\frac{\partial}{\partial\mu} + \left(- \frac{a_0}{a_0^\prime} \epsilon + \beta\right)\frac{\partial}{\partial g} + \left(\left(1 + \frac{a_0}{a_0^\prime}\,\frac{d_0^\prime}{d_0}\right)\epsilon
+ \gamma\right) A\frac{\partial}{\partial A}\right]A_B.\eqno(61)$$
From eqs. (59) and (61) we obtain
$$-\frac{a_0}{a_0^\prime} d^\prime_{n+1} + \beta d_n^\prime + \frac{a_0}{a_0^\prime} \,\frac{d_0^\prime}{d_0} d_{n+1} + \gamma d_n = 0.\eqno(62)$$
When $n = 0$, eq. (62) serves to fix $\gamma$ in terms of $d_0$, $d_1$, $a_0$ and $a_1$. Just as eq. (51) leads to eq. (56), so also eq. (62) leads to the sum of all first order poles in eq. (59) being given by 
$$R_0(U) = (1 - B_3 U)^{-D_2/2B_3}.\eqno(63)$$
(We have made the expansion
$$\gamma (g) = \sum_{n=0}^\infty D_{2n}g^{2n}\;.)\eqno(64)$$
Also, just as eq. (58) was derived, we find that
$$A_B = \mu^{-\epsilon} A\exp - \int_0^g dx\left[\left(\frac{a_0}{a_0^\prime}\,\frac{d_0^\prime}{d_0} \epsilon + \gamma\right)/\left(-\frac{a_0}{a_0^\prime} \epsilon + \beta\right)\right].\eqno(65)$$

If $g$ and $A$ are the gauge coupling and Yang-Mills field respectively, then gauge invariance implies that [10,11]
$$g_B A_B = gA.\eqno(66)$$
From eqs. (1) and (59), eq. (66) implies that
$$g = \left( \sum_{n=0}^\infty\,\frac{a_n(g)}{\epsilon^n}\right)\left(\sum_{m=0}^\infty\,\frac{d_m(g)}{\epsilon^m}\right)\eqno(67)$$
so that
$$a_0 d_0 = g\eqno(68)$$
$$a_1 d_0 + a_0 d_1 = 0\eqno(69)$$
$$a_2 d_0 + a_1d_1 + a_0d_2 \eqno(70)$$
or, in general
$$\sum_{k=0}^\infty a_{n-k} d_k = 0\;\;\;\; ( n > 0).\eqno(71)$$
By eqs. (68) and (69)
$$d_0 = g/a_0\eqno(72)$$
$$d_1 = -a_1 d_0/a_0\eqno(73)$$
which can be used with eqs. (11) and (62) to show [10,11]
$$\beta (g) = -g\gamma(g).\eqno(74)$$
Because of eqs. (72-74), eqs. (31) and (65) are consistent with eq. (66). It also follows from eqs. (32) and (66) that
$$\lim_{\epsilon \rightarrow 0} A_B = \infty \;.\eqno(75)$$
We now examine formal solutions of the RG equation.
\section{Formal Solution of the RG Equation}

We now consider the formal solution of the RG equation, using the particular example of the RG equation for the effective potential in a $\lambda \phi_4^4$ model with no Lagrangian mass term.  The RG equation is
$$\left( \mu \frac{\partial}{\partial\mu} + \beta (\lambda)\frac{\partial}{\partial\lambda} + \gamma (\lambda) \phi \frac{\partial}{\partial\phi}\right) V(\mu , \lambda , \phi) = 0\eqno(76)$$
where $\lambda = \lambda(\mu)$ and $\phi = \phi (\mu)$ satisfy
$$\mu \frac{d\lambda(\mu)}{d\mu} = \beta (\lambda(\mu))\eqno(77)$$
$$\mu \frac{d\phi(\mu)}{d\mu} = \gamma (\lambda(\mu))\phi(\mu).\eqno(78)$$
(The dependence of $\lambda$ and $\phi$ on $\mu$ will from now on will not be written explicitly.) Often the solution to eq. (76) is found by first taking
$$V(\mu, \lambda, \phi) = Y(\lambda, L)\phi^4\eqno(79)$$
where
$$L = \ln (\phi /\mu).\eqno(80)$$
Eq. (76) then becomes
$$\left[ \frac{\partial}{\partial L} + \tilde{\beta} (\lambda) \frac{\partial}{\partial \lambda} + 4\tilde{\gamma} (\lambda)\right] Y(\lambda , L) = 0\eqno(81)$$
where
$$\tilde{\beta} = \beta /(-1+\gamma)\eqno(82)$$
$$\tilde{\gamma} = \gamma/(-1 + \gamma ).\eqno(83)$$
An auxiliary function $\overline{\lambda} (L,\lambda)$ is now defined by the integral
$$L = - \int_\lambda^{\overline{\lambda}(L,\lambda)} \,\frac{dx}{\tilde{\beta}(x)}\eqno(84)$$
so that
$$\overline{\lambda}(L = 0, \lambda) = \lambda\eqno(85)$$
and
$$\frac{\partial\overline{\lambda}(L,\lambda )}{\partial L} = -\tilde{\beta} (\overline{\lambda} (L,\lambda))\eqno(86)$$
$$\frac{\partial\overline{\lambda}(L,\lambda )}{\partial \lambda} = +
\frac{\tilde{\beta} (\overline{\lambda} (L,\lambda))}{\tilde{\beta}(\lambda)}\,.\eqno(87)$$
The solution to eq. (81) is often taken to be [12-15]
$$Y(\lambda , L) = f(\overline{\lambda} (L, \lambda))\exp\left[ -4 \int_0^L \tilde{\gamma}(\overline{\lambda}(x,\lambda))dx\right].\eqno(88)$$
However, substitution of eq. (88) into eq. (81) leads to
$$\left\lbrace f^\prime(\overline{\lambda} (L, \lambda))
\left(\frac{\partial\overline{\lambda}(L,\lambda )}{\partial L} + \tilde{\beta} (\lambda)\frac{\partial\overline{\lambda}(L,\lambda )}{\partial \lambda}\right)\right.\eqno(89)$$
$$+ f(\overline{\lambda} (L, \lambda))
\left(-4\tilde{\gamma}(\overline{\lambda}(L,\lambda)) -4 \int_0^L 
\frac{\partial\overline{\gamma}(\overline{\lambda}(x,\lambda))}{\partial\lambda} dx\right. \nonumber$$
$$ \left. + 4 \tilde{\gamma}(\lambda)\right)\Bigg\} \exp\left[-4\int_0^L \tilde{\gamma}(\overline{\lambda}(x,L))dx\right] = 0.\nonumber$$
By eqs. (86) and (87), the term on the left side of eq. (89) proportional to $f^\prime$ vanishes; the remaining term though is non-zero if $\tilde{\gamma} \neq 0$ except at the point $L = 0$ (on account of eq. (85)).

A formal solution to eq. (81) can be found by following the approach of ref. [16] . Perturbative calculations show that the form of the solution to eq. (81) is
$$Y(\lambda , L) = \sum_{m=1}^\infty \sum_{n=0}^{m-1} a_{mn} \lambda^m L^n\eqno(90)$$
$$\equiv \sum_{n=0}^\infty A_n (\lambda) L^n.\eqno(91)$$
From eq. (81) it then follows that
$$A_{n+1} (\lambda) = \frac{-1}{n+1} \left[\tilde{\beta}(\lambda)\frac{\partial}{\partial\lambda} + 4 \tilde{\gamma}(\lambda)\right]\tilde{A}_n (\lambda).\eqno(92)$$
If now
$$A_n(\lambda) = \left[\exp - 4 \int_{\lambda_{0}}^\lambda dx \frac{\tilde{\gamma}(x)}{\tilde{\beta}(x)}\right]\tilde{A}_n (\lambda).\eqno(93)$$
then eq. (92) becomes
$$\tilde{A}_{n+1} (\lambda) = \frac{-1}{n+1} \tilde{\beta}(\lambda)\frac{\partial}{\partial\lambda}\tilde{A}_n (\lambda),\eqno(94)$$
or
$$\tilde{A}_{n+1} (\eta) = \frac{1}{n+1} \frac{\partial}{\partial\eta}\tilde{A}_n (\eta)\eqno(95)$$
where
$$\eta(\lambda) = -\int_{\lambda_{0}}^\lambda \frac{dx}{\tilde{\beta}(x)}.\eqno(96)$$
We then have by eqs. (90)-(96)
$$Y(\lambda , L) = \exp\left[-4 \int_{\lambda_{0}}^\lambda dx \frac{\tilde{\gamma}(x)}
{\tilde{\beta}(x)}\right]\sum_{n=0}^\infty \tilde{A}_n (\lambda) L^n\nonumber$$
$$= \exp\left[-4 \int_{\lambda_{0}}^\lambda dx \frac{\tilde{\gamma}(x)}
{\tilde{\beta}(x)}\right]\sum_{n=0}^\infty \frac{L^n}{n!} \left(\frac{d}{d\eta}\right)^n \tilde{A}_0 (\lambda(\eta)) \nonumber$$
$$= \exp\left[-4 \int_{\lambda_{0}}^\lambda dx \frac{\tilde{\gamma}(x)}
{\tilde{\beta}(x)}\right]\tilde{A}_0 (\lambda(\eta + L)) \nonumber$$
$$= \exp\left[-4 \int_{\overline{\lambda}(\eta + L)}^\lambda dx \frac{\tilde{\gamma}(x)}
{\tilde{\beta}(x)}\right]A_0 (\lambda(\eta + L)).\eqno(97)$$
Eq. (97) is distinct from eq. (88).

The function $A_0(\lambda(\eta + L))$ can be determined by applying renormalization conditions to the effective potential $V$. (The functions $\beta$ and $\gamma$ are themselves affected by these renormalization conditions.) For example, if we impose the renormalization condition [15]
$$V = \frac{\lambda\phi^4}{4!}\eqno(98)$$
when $\phi (\mu) = \mu$ (so that $L = 0$), then since $\lambda(\eta) = \lambda$ eq. (79),
$$A_0(x) = \frac{x}{4!}\eqno(99)$$
reducing eq. (97) to
$$V(\mu , \lambda , \phi) = \exp\left[- \int^\lambda_{\lambda_{(\eta + L)}} dx \frac{\tilde{\gamma}(x)}
{\tilde{\beta}(x)}\right]
\frac{\lambda(\eta + L)}{4!}\phi^4.\eqno(100)$$
Other renormalization conditions such as [12]
$$\frac{d^4V}{d\phi^4}\left|_{\phi = \mu} = \lambda\right.\eqno(101)$$
lead to a differential equation for $A_0(\lambda)$.

We now turn to solving the RG equations by the method of characteristics.

\section{The Method of Characteristics}

The method of characteristics is discussed in conjunction with the RG equation in refs. [17-20]. We persue this approach to solving the RG equation by first examining a toy model.

Let us consider the first order linear homogeneous partial differential equation
$$x \frac{\partial A(x,y)}{\partial x} + y^2 \frac{\partial A(x,y)}{\partial y} = 0.\eqno(102)$$
A solution to this equation is
$$A_0(x,y) = xe^{1/y}.\eqno(103)$$
If we define ``characteristic functions'' $\overline{x}(t)$ and $\overline{y}(t)$, 
using
$$\frac{d\overline{x}(t)}{dt} = \overline{x}(t)\;\;\;\;\; (\overline{x}(0) = x)\eqno(104)$$
$$\frac{d\overline{y}(t)}{dt} = \overline{y}^2(t)\;\;\;\;\; (\overline{y}(0) = y)\eqno(105)$$
then
$$\frac{dA_0(\overline{x}(t), \overline{y}(t))}{dt} = \overline{x}(t) \
\frac{\partial A_0(\overline{x}(t),\overline{y}(t))}
{\partial \overline{x}(t)}
+ \overline{y}^2(t)
\frac{\partial A_0(\overline{x}(t),\overline{y}(t))}
{\partial \overline{y}(t)} = 0\eqno(106)$$
and hence $A_0(\overline{x}(t), \overline{y}(t)$) is independent of $t$. As a result, it follows that
$$x\frac{\partial}{\partial x} A_0 (\overline{x}(t), \overline{y}(t)) + y^2 \frac{\partial}{\partial y} A_0 (\overline{x}(t), \overline{y}(t)) = 0\eqno(107)$$
so that $A_0(\overline{x}(t), \overline{y}(t)$) is also a solution to eq.(102). To be explicit, eqs. (104) and (104) have solutions
$$\overline{x}(t) = xe^t\eqno(108)$$
$$\overline{y}(t) = \frac{y}{1 - yt}\eqno(109)$$
Substitution of eqs. (108) and (109) into eq. (103) shows that for all $t$
$$A_0(\overline{x}(t), \overline{y}(t)) = A_0(x,y)\eqno(110)$$
identically, so that eq. (106) is indeed satisfied.

We also note that the method of characteristics can also be applied to the equation
$$f(x,y)\frac{\partial A(x,y)}{\partial x} + g(x,y) \frac{\partial A(x,y)}{\partial y} + h(x,y) A(x,y) = 0.\nonumber$$
If $A_0(x,y)$ is a solution to this equation and eqs. (104) and (105) define $\overline{x}(t)$ and $\overline{y}(t)$, then
$$A(\overline{x}(t)), \overline{y}(t), t) = A_0(\overline{x}(t), \overline{y}(t), (t))\exp \int_0^t h(\overline{x}(t^\prime), \overline{y}(t^\prime))dt^\prime\nonumber$$
reduces to $A_0(x,y)$ when $t = 0$ and $\frac{d}{dt} A(\overline{x}(t), \overline{y}(t), t) = 0$.

Suppose now we were unable to find the exact solution of eq. (103) but were only able to establish the approximate solution
$$A_0^{(1)} (x,y) = x\left(1 + \frac{1}{y}\right).\eqno(111)$$
One way of generating corrections to $A_0^{(1)} (x,y)$ would be to make the ansatz
$$A_0 (x,y) = x \sum_{n = 0}^\infty \alpha_n\left(\frac{1}{y}\right)^n\eqno(112)$$
with $\alpha_0 = \alpha_1 = 1$. Substituting eq. (112) into eq. (102) yields the recursion relation
$$\alpha_n = (n+1)\alpha_{n+1},\eqno(113)$$
allowing us to recover the solution of eq. (103) just knowing $\alpha_0 = 1$. This is analogous to preforming the ``sum of logarithms (poles)'' in refs. [1-3,8].

The constants $\alpha_n$ in eq. (112) can also be found by substitution of the characteristic functions of eqs. (108) and (109) into eq. (112) so that by eq. (110),
$$x\sum_{n=0}^\infty \alpha_n \left(\frac{1}{y}\right)^n = xe^t \sum_{n=0}^\infty \alpha_n \left(\frac{1-yt}{y}\right)^n.\eqno(114)$$
This is true for all $t$ only if eq. (113) is satisfied.

An alternate to making the ansatz of eq. (112) in order to determine the exact solution of eq. (103) from the approximate solution of eq. (111) is to substitute the characteristic functions to eqs. (108) and (109) into the approximate solution of eq. (111) so that
$$A_0^{(1)} (\overline{x}(t), \overline{y}(t)) = xe^t\left[1 + \frac{1-yt}{y}\right].\eqno(115)$$
Unlike $A_0 (\overline{x}(t), \overline{y}(t))$, the $t$ dependence in $A_0^{(1)} (\overline{x}(t), \overline{y}(t))$ does not cancel out. However, at a particular value of $t$ (namely $t = \frac{1}{y}$)
$$A_0^{(1)} \left(\overline{x}\left(t = \frac{1}{y}\right), \overline{y}\left(t = \frac{1}{y}\right)\right) = xe^{\frac{1}{y}} = A_0 (x,y)\eqno(116)$$
so that the exact solution is recovered.

The occurence of a particular value of the characteristic parameter $t$ at which the perturbative approximation to a differential equation evaluated at the characteristic functions reduces to the exact solution is a mirror of the approach outlined in ref. [18] to the ``summation of leading logs''.

There are two non-trivial complications that occur in the RG equation which arises in perturbative quantum field theory that distinguish it from our model.  First of all, the functions $\beta(g)$ and $\gamma(g)$ occurring in, say, eq. (76) are not known exactly so one is able to only determine the sum of leading log contributrions to the exact solution to the RG equations if $\beta$ and $\gamma$ are known to one-loop order etc.

The second point that distinguishes the RG equation from the toy model of eq. (102) is that in the RG equation the coupling and background fields ($\gamma$ and $\phi$ respectively in eq. (76)) are dependent on the scale perimeter $\mu$ (as indicated in eqs. (78) and (79)). By way of contrast, in eq. (102) $x$ and $y$ are independent. However, in actually applying the method of characteristics to analyze the RG equation, this dependence of $g$ and $\phi$ on $\mu$ is immaterial. It should be noted though that the form of the equations for the ``running coupling'' $\lambda(\mu)$
$$\mu \frac{d\lambda(\mu)}{d\mu} = \beta(\lambda(\mu))\;\;\;\;\;(\lambda(\Lambda) = \infty)\eqno(116)$$
is identical in form to the equation for the characteristic function
$$\frac{d\overline{\lambda}(t)}{dt} = \beta(\overline{\lambda}(t))\;\;\;\;(\overline{\lambda}(0) = \lambda(\mu)).\eqno(117)$$
In eq. (117), the running coupling serves as a boundary condition to the characteristic coupling. The perturbative approaches used to solve the running and characteristic functions are quite distinct.

We now illustrate how the method of characteristics can be used in conjunction with the RG equation by considering the free energy in thermal QCD.

\section{The Free Energy in Thermal QCD}

The free energy in QCD at high $T$ in general has the form 
$$F/F_0 = 1 + \sum_{n=0}^\infty \sum_{m=0}^\infty\left\lbrace A_{n+m,m} x
+ B_{n+m,m}x^{3/2} + C_{n+m,m} x^2\ln x\right\rbrace x^{m+n} L^m\eqno(118)$$
where $x$ is the QCD coupling and $L = \ln \left(\mu^2/(2\pi T)^2\right)$.
In ref. [21], the RG equation
$$\left( \mu^2 \frac{\partial}{\partial\mu^2} + \beta (x(\mu))\frac{\partial}{\partial x(\mu)}\right)\left(F/F_0\right) = 0\eqno(119)$$
where
$$\mu^2 \frac{dx(\mu)}{d\mu^2} = \beta(x(\mu)) = b_2 x^2(\mu) + b_3 x^3(\mu) + \ldots\eqno(120)$$
is used to generate ordinary differential equations for the functions
$$R_n (U) = \sum_{m=0}^\infty  A_{n+m,m} U^m,\;\; S_n (U) = \sum_{m=0}^\infty  B_{n+m,m} U^m,\;\;T_n (U)= \sum_{m=0}^\infty  A_{n+m,m} U^m.\eqno(121)$$
These equations are solved subject to the boundary conditions 
$R_n(0) = A_{n,0}$, $S_n(0) = B_{n,0}$, $T_n(0) = C_{n,0}$.

We can also approach the problem of obtaining these higher order corrections to the free energy by using the method of characteristics, much as was done in ref. [18].  A parameter $\hbar$ is introduced to organize the evaluation of $F/F_0$ into a systematic perturbative expansion.  We begin by rescaling the characteristic parameter $t$ and the characteristic function $\overline{x}(t)$
$$t \rightarrow t/\hbar \eqno(122)$$
$$\overline{x} (t) = \hbar \overline{x} (t)\eqno(123)$$
so that
$$\hbar \frac{d\overline{\mu}^2(t)}{dt} = \overline{\mu}^2(t) \;\;\left(\overline{\mu}^2(0) = \mu^2\right)\eqno(124)$$
$$\hbar^2 \frac{d\overline{x}(t)}{dt} = \beta(\hbar \overline{x}(t))\;\;\;\;\;(\overline{x}(0) = x)\nonumber$$
$$\;\;\;\;\;\;\;\;\;\;\;\;\;\;\;\;\;\;\;\;\;= b_2 \hbar^2 \overline{x}^2(t) + b_3 \hbar^3 \overline{x}^3(t) + \ldots .\eqno(125)$$
If now $\overline{x}(t)$ is itself expanded in powers of $\hbar$
$$\overline{x}(t) = \sum_{n=0}^\infty \hbar^n \overline{x}_n(t)\eqno(126)$$
with
$$\overline{x}_n(0) = x(\mu)\delta_{n0}\eqno(127)$$
we find that
$$\frac{d}{dt}\left(\overline{x}_0(t) + \hbar \overline{x}_1 (t) + \ldots\right)
= \left(b_2 \overline{x}_0^2 (t) \right) + \hbar \left(b_3 \overline{x}_0^3 (t) + 2b_2 \overline{x}_0 (t) \overline{x}_1(t)\right) + \ldots .\eqno(128)$$
By equating like powers of $\hbar$ in eq. (128), a sequence of equations for $\overline{x}_0(t)$, $\overline{x}_1(t) \ldots$ are found. (This is contingent on knowing $b_2$, $b_3 \ldots$ etc.). Once these are known, these can be substituted into the expansion of eq. (118) for $F/F_0$.
Using 
$$\ln\left(\overline{x}_0 + \hbar \overline{x}_1 + \ldots\right) = \ln \overline{x}_0 + \left(\frac{\hbar \overline{x}_1  + \hbar^2 \overline{x}_2 + \ldots}{\overline{x}_0}\right)
-\frac{1}{2} \left(\frac{\hbar \overline{x}_1  + \hbar^2 \overline{x}_2 + \ldots}{\overline{x}_0}\right)^2 + \ldots \eqno(129)$$
we find that eq. (118) becomes
$$F/F_0 \left(\overline{x}(t), \overline{\mu}^2 (t)\right) = 1 + \hbar\left(A_{00} \overline{x}_0\right) + \hbar^2\left(A_{00}\overline{x}_1
+ A_{11} \overline{x}_0^2 \overline{L} + C_{00} \overline{x}_0^2 \ln \overline{x}_0 + A_{10}\overline{x}_0^2\right)\nonumber$$
$$+ \hbar \ln \hbar \left(C_{00} \overline{x}^2_0\right) + \hbar^{3/2} \left(B_{00} \overline{x}_0^{3/2}\right)
+ \hbar^{5/2} \left(\frac{3}{2} B_{00} \overline{x}_0^{1/2} \overline{x}_1 + B_{11} \overline{x}_0^{5/2}\overline{L} + B_{10} \overline{x}_0^{5/2}\right) + \ldots \eqno(130)$$
where $\overline{L} = \ln\left(\overline{\mu}^2(t)/(2\pi T)^2\right)$. From eqn. (128) we find that
$$\overline{\mu}^2(t) = \mu^2 e^t\eqno(131)$$
$$\overline{x}_0 (t) = \frac{x(\mu)}{1 - b_2 x(\mu)t}\eqno(132)$$
$$\overline{x}_1(t) = -\frac{b_3}{b_2} \,\frac{x^2(\mu)\ln\left(1 - b_2 x(\mu)t\right)}{\left(1 - b_2 x(\mu)t\right)^2}.\eqno(133)$$
Eqs. (131-133) now can be substituted into eq. (130) and we can eventually set $\hbar = 1$.

Different values of $t$ can now be considered. If $t = 0$ in eqs. (130-133) then eq. (118) is recovered. If instead $t = \hbar \ln k$, in eqs. (130-133), then eq. (130) would differ from eq. (118) by factors of $\ln k$. Since $F/F_0 (\overline{x}(t),\overline{y}(t))$ is independent of $t$ (and hence of $\ln k$), setting coefficients of powers of $\ln k$ equal to zero provides sets of consistency conditions that fix $A_{p,q}$, $B_{p,q}$, $C_{p,q}$ $(q > 0)$ in terms of 
$A_{p,0}$, $B_{p,0}$, $C_{p,0}$.

The most useful value of $t$ is the one for which $\overline{L} = 0$
$$t = \hbar \ln \left((2\pi T)^2/\mu^2\right).\eqno(134)$$
With this value of $t$, then to $O(\hbar^{5/2})$
$$F/F_0 = 1 + \left[ A_{00} \frac{x}{w}\right] + \left[A_{00}\left(\frac{-rx^2\ln w}{w^2}\right) + A_{10}\left(\frac{x^2}{w^2}\right)\right.\nonumber$$
$$\left. + C_{00}\left(\frac{x^2}{w^2} \ln \frac{x}{w}\right)\right] + \left[ B_{00} \left(\frac{x}{w}\right)^{3/2}\right]\eqno(135)$$
$$+\left[B_{00}\left(-\frac{3}{2} r\frac{x^{5/2}}{w^{1/2}} \left(\frac{x^2\ln w}{w^2}\right)\right) + B_{10}\left(\frac{x}{w}\right)^{5/2}\right]\nonumber$$
where $w = 1 + b_2 x L$ and $r = b_3/b_2$. This is identical to what is obtained in ref. [21] by solving the RG equation (119) for the functions $R_0$, $S_0$, $T_0$, $R_1$ and $T_1$ of eq. (121). It is reasonable to assume that this agreement persists beyond the order in perturbation that we have considered.

We now turn to applying the method of characteristics to the relationship between $g$ and $g_B$ discussed in ref. [8] and section 3 above.

\section{Relating $g$ and $g_B$}

Eqs. (8) and (9) together with eq. (12) imply that
$$\mu \frac{\partial g_B}{\partial \mu} + \left(-\epsilon g + \beta(g)\right)\frac{\partial g_B}{\partial g} = 0\eqno(136)$$
in the MS scheme. In ref. [8], the functions $S_n$ of eqs. (5-7) are determined by solving a series of nested ordinary differential equations arising from the RG equation. It is of interest to presently apply the method of characteristics to eq. (136).

Characteristic functions are first introduced satisfying
$$\frac{d\overline{\mu}(t)}{dt} = \overline{\mu} (t)\;\;\;\;\; (\overline{\mu} (0) = \mu)\eqno(137)$$
$$\frac{d\overline{g}(t)}{dt} = -\epsilon \overline{g}(t) + \beta(\overline{g}(t)) \;\;\;\;\; (\overline{g} (0) = g).\eqno(138)$$
As in the previous section, a scaling parameter $\hbar$ is introduced to organize our perturbative expansion. We begin by making the rescalings
$$\epsilon \rightarrow \epsilon\hbar ,\;\;\;t \rightarrow t/\hbar \;\;\;\;\;\overline{g}(t)\rightarrow \overline{g} (t)\hbar^{1/2}\eqno(139)$$
and the expanding
$$\overline{g}(t) = \overline{g}_0(t) + \hbar\overline{g}_1(t) +\hbar^2            \overline{g}_2(t) + \ldots\eqno(140)$$
with
$$\overline{g}(0) = g\delta_{n,0}.\eqno(141)$$
From eq. (137)
$$\overline{\mu}(t) = \mu e^{t/k}\eqno(142)$$
and from eq. (138-141)
$$\overline{g}_0^2 = \frac{g^2}{\frac{b_3}{\epsilon} g^2\left(1 - e^{2\epsilon t}\right) + e^{2\epsilon t}}\eqno(143)$$
and
$$\overline{g}_1^2 = \frac{b_5 e^{2\epsilon t}g^5}{2\epsilon\left[\frac{b_3g^2}{\epsilon} + \left(1 - \frac{b_3g^2}{\epsilon}\right)e^{2\epsilon t}\right]^{3/2}}
\left\lbrace \frac{1 - \frac{b_3}{\epsilon} g^2}{\left(\frac{b_3g^2}{\epsilon}\right)^2}\right.\nonumber$$
$$\left. \ln \left(
\frac{
\frac{b_3g^2}{\epsilon} + \left(1 - \frac{b_3g^2}{\epsilon}\right)e^{2\epsilon t}}
{e^{2\epsilon t}}
\right) + \frac{1}{\left(\frac{b_3g^2}{\epsilon}\right)}\left(1 - \frac{1}{e^{2\epsilon t}}\right)\right\rbrace .\eqno(144)$$

We now note that if eqs. (139) and (140) are used in conjunction with eq. (1) and (3)
$$g_B\left(\overline{\mu}(t), \overline{g}(t)\right) = \overline{\mu}^{\epsilon\hbar}\left\lbrace\hbar^{1/2}\left[\overline{g}_0 + \frac{a_{11}\overline{g}_0^3}{\epsilon} + \frac{a_{22}\overline{g}_0^5}{\epsilon^2} + \ldots\right]\right.\eqno(145)$$
$$\left. +\hbar^{3/2} \left[\left(\overline{g}, + \frac{3a_{11}\overline{g}_0^2\overline{g}_1}{\epsilon} +
\frac{5a_{22}\overline{g}_0^4\overline{g}_1}{\epsilon^2} + \ldots\right)
+\left(\frac{a_{21}\overline{g}_0^5}{\epsilon} +
\frac{a_{31}\overline{g}_0^7}{\epsilon^2} + \ldots\right)\right] + O(\hbar^{5/2})\right\rbrace .\nonumber$$
In the limit $t \rightarrow \infty$, by eqs. (143 - 144)
$$\overline{g}_0 \rightarrow \frac{g}{\sqrt{1 - \frac{b_3g^2}{\epsilon}}} e^{-\epsilon t}\eqno(146)$$
and
$$\overline{g}_1 \rightarrow \frac{b_5 g^5}{2\epsilon} 
\frac{1}{\left(1 - \frac{b_3g^2}{\epsilon}\right)^{3/2}}
\left\lbrace \frac{1 - \frac{b_3g^2}{\epsilon}}{\left(\frac{b_3g^2}{\epsilon}\right)^2}
\ln\left(1 - \frac{b_3g^2}{\epsilon}\right) + \frac{1}{\left(\frac{b_3g^2}{\epsilon}\right)}\right\rbrace e^{-\epsilon t}\eqno(147)$$
and
$$\overline{\mu}^{\epsilon \hbar} \rightarrow \mu^{\epsilon\hbar} e^{\epsilon t}.\eqno(148)$$
Together, eqs. (146-148) reduce eq. (145) in the limit $t \rightarrow \infty$ to
$$g_B = \mu^{\epsilon\hbar}\left\lbrace\hbar^{1/2} \left[\frac{g}{\sqrt{1-\frac{b_3g^2}{\epsilon}}}\right] + \hbar^{3/2}\left[\frac{b_5g^5}{2\epsilon}\,\frac{1}{\left(1 - \frac{b_3g^2}{\epsilon}\right)^{3/2}}\right.\right.\nonumber$$
$$ \left. \left. \left( \frac{1 - \frac{b_3g^2}{\epsilon}}{\left(\frac{b_3g^2}{\epsilon}\right)^2} \ln \left(1 - \frac{b_3g^2}{\epsilon}\right) + \frac{1}{\left(\frac{b_3g^2}{\epsilon}\right)}\right)\right] + \ldots \right\rbrace\eqno(149)$$
as terms proportional to powers of $\overline{g}_i$ greater than one vanish.  In the limit $\hbar \rightarrow 1$, eq. (149) reduces to what is obtained if the function $S_0$ and $S_1$ computed in ref. [8] are substituted into eq. (5). It is reasonable to anticipate that the method of characteristics can be used to reproduce all of the terms $S_n$ in eq. (5) upon letting $t \rightarrow \infty$.

The application of the method of characteristics to the RG equation for the effective potential $V$ in massless scalar electrodycnamics ($\not\!\!M$ SQED) will now be discussed.

\section{ V in $\not\!\!M$ SQED}

The effective potential $V$[12, 22-24] is a tool for analyzing spontaneous symmetry breaking. The possibility that when RG accessible radiative effects are included in $V$, realistic spontaneous symmetry breaking patterns could occur, even if no bare mass term is in the classical action, has been explored in the context of $\not\!\!M$ SQED and the massless standard model in refs. [3, 25].

In order to determine the LL contributions to $V$ in refs. [3, 25], these contributions were themselves expanded in powers of the gauge coupling in both of the models considered. The need for making this expansion can be overcome by making use of the method of characteristics provided one can solve for the appropriate characteristic functions.

The model being considered is $\not\!\!M$ SQED where [12]
$$L = \frac{1}{2}\left(\partial_\mu \phi_1 - eA_\mu \phi_2\right)^2 + \frac{1}{2} \left(\partial_\mu \phi_2 + eA_\mu\phi_1\right)^2 - \frac{\lambda}{4!} \left(\phi_1^2 + \phi_2^2\right)^2 ,\eqno(150)$$
where $A_\mu$ is a $U(1)$ gauge field and $(\phi_1,\phi_2)$ are scalars. The one-loop effective potential in the presence of a background $\phi^2 = \phi_1^2 + \phi^2_2$ is [12]
$$V^{(1)} (\phi) = \left[ \frac{\lambda}{4!} + \left( \frac{5\lambda^2}{1152\pi^2} + \frac{3e^4}{64\pi^2} \right) \left(\ln \,\frac{\phi^2}{\mu^2} + K\right)\right]\phi^4.\eqno(151)$$
The constant $K$ is determined by the renormalization condition used; in ref. [12] the requirement
$$\frac{d^4V}{d\phi^4} \,(\phi = \mu) = \lambda\eqno(152)$$
fixes
$$K = -\frac{25}{6}.\eqno(153)$$
At one-loop order the RG functions are 
$$\frac{\mu}{\phi}\,\frac{d\phi}{d\mu} = \gamma = \frac{3e^3}{16\pi^2}\eqno(154)$$
$$\mu \frac{de^2}{d\mu} = \beta_e = \frac{e^4}{24\pi^2}\eqno(155)$$
$$\mu \frac{d\lambda}{d\mu} = \beta_\lambda = \frac{5}{24\pi^2} \lambda^2 - \frac{3}{4\pi^2} \lambda e^2 + \frac{9}{4\pi^2} e^4. \eqno(156)$$
(The dependence of $\phi$, $e^2$ and $\lambda$ on $\mu$ is not noted explicitly.) As is discussed in section 3, these one-loop results do not depend on having used the renormalization condition of eq. (152).

We now employ the method of characteristics to obtain the RG accessible higher order effects that follow form eqs. (151) and (154-156). If now we take $x = e^2/4\pi$ and $y = \lambda/4\pi^2$, we write
$$V = \sum_{m=1}^\infty \sum_{n=1}^\infty \sum_{k=0}^{m+n-1} \,a_{mnk} x^m y^n L^k \phi^4\eqno(157)$$
where, by eq. (151), $a_{010} = \pi^2/6$, $a_{021} = 5\pi^2/72 = (-6/25)a_{020}$, $a_{201} = 3/4\pi^2 = (-6/25)a_{200}$. The RG functions in general are of the form
$$\frac{\mu}{\phi}\,\frac{d\phi}{d\mu} = \gamma = \sum_{m,n=1}^\infty\,p_{mn} x^my^n\eqno(158)$$
$$\mu\,\frac{dx}{d\mu} = \beta_x = \sum_{m,n=1}^\infty\,q_{mn} x^my^n\eqno(159)$$
$$\mu\,\frac{dy}{d\mu} = \beta_y = \sum_{m,n=1}^\infty\,r_{mn} x^my^n\eqno(160)$$
and the RG equation is
$$\left(\mu \frac{\partial}{\partial \mu} + \beta_x \frac{\partial}{\partial x} + \beta_y \frac{\partial}{\partial y} + \gamma\phi \frac{\partial}{\partial \phi}\right) V(\mu, x(\mu), y(\mu), \phi(\mu)) = 0.\eqno(161)$$
As usual, in applying the method of characteristics to eq. (161) involve the characteristic functions $\overline{\mu}(t)$, $\overline{x}(t)$, $\overline{y}(t)$, $\overline{\phi}(t)$
$$\frac{d\overline{\mu}(t)}{dt} = \overline{\mu}(t)\;\;\;\;\; (\overline{\mu}(0) = \mu)\eqno(162)$$
$$\frac{d\overline{x}(t)}{dt} = \beta_x\left(\overline{x}(t), \overline{y}(t)\right)\;\;\;\;\; (\overline{x}(0) = x(\mu))\eqno(163)$$
$$\frac{d\overline{y}(t)}{dt} = \beta_y\left(\overline{x}(t), \overline{y}(t)\right)\;\;\;\;\; (\overline{y}(0) = y(\mu))\eqno(164)$$
$$\frac{1}{\overline{\phi}(t)}\,\frac{d\overline{\phi}(t)}{dt} = \gamma\left(\overline{x}(t), \overline{y}(t)\right)\;\;\;\;\; (\overline{\phi}(0) = \phi(\mu)).\eqno(165)$$
As in the two previous sections, we introduce an order-counting parameter $\hbar$ so that $t \rightarrow t/\hbar$, $\overline{x} \rightarrow \overline{x}\hbar$, $\overline{y} \rightarrow \overline{y} \hbar$, $\overline{\phi} \rightarrow \overline{\phi}$ and make the expansions $\overline{x} = \displaystyle{\sum_{n = 0}^\infty} \hbar^n \overline{x}_n$ ($\overline{x}_n (0) = x(\mu)\delta_{n0}$) and similarly for $\overline{y}$ and $\overline{\phi}$. To lowest order in $\hbar$, by eqs. (151) and (154-156),
$$V = \frac{\hbar^2}{6} \overline{y}_0(t) \overline{\phi}_0(t) \eqno(166)$$
$$\frac{d\overline{x}_0}{dt} = \frac{1}{6} \overline{x}_0\eqno(167)$$
$$\frac{d\overline{y}_0}{dt} = \frac{5}{6} \overline{y}_0^2 = 3\overline{x}_0\overline{y}_0 + 9\overline{x}_0^2\eqno(168)$$
$$\frac{d\overline{\phi}_0}{dt} = \frac{3}{4} \overline{x}_0 \overline{\phi}_0.\eqno(169)$$
From eq. (167) we obtain
$$\overline{x}_0 = \frac{x}{1 - \frac{1}{6} xt}\eqno(170)$$
and thus eq. (169) gives
$$\overline{\phi}_0 = \frac{\phi}{(1 - \frac{1}{6} xt)^{9/2}}\,.\eqno(171)$$
Eq. (168) is a Ricotti equation, but instead of solving it directly, we follow ref. [12] and use eqs. (167) and (168) to obtain
$$\frac{d\overline{y}_0}{d\overline{x}_0} = 5\left(\frac{\overline{y}_0}{\overline{x}_0}\right)^2 - 18 \left(\frac{\overline{y}_0}{\overline{x}_0}\right) + 54, \eqno(172)$$
whose solution is
$$\overline{y}_0 = \overline{x}_0\left\lbrace 
\frac{\sqrt{719}}{10} \tan \left[\frac{\sqrt{719}}{2}
\ln \left(\overline{x}_0 \not\!\!{C}\right)\right] + \frac{19}{10}\right\rbrace\eqno(173)$$
where $\not\!\!{C}$ is fixed by the condition $\overline{y}_0 (0) = y(\mu)$. Eqs. (170), (171) and (173) can now be substituted into eq. (166) to obtain the RG improved effective potential, as determined by the method of characteristics.

It is now necessary to fix the parameter $t$. The analogue of eq. (134) would be to choose $t$ so that
$\overline{L} = \ln \left(\overline{\phi}(t)/\overline{\mu}(t)\right) = 0$, but this would result in an awkward transcendental equation for $t$. If instead we followed the suggestion of refs. [19, 20] selecting
$$t = \ln (\phi/\mu)\eqno(174)$$
then $V$ becomes
$$V = \frac{\pi^2}{6} \frac{x}{w}\left\lbrace \frac{\sqrt{719}}{10}\tan\left[\frac{\sqrt{719}}{2}\ln \left(\frac{x}{w} \not\!\!{C}\right)\right] + \frac{19}{10}\right\rbrace\left(\frac{\phi}{w^{9/12}}\right)^2\eqno(175)$$
where $w = 1 - \frac{1}{6} x\ln\frac{\phi}{\mu}$. A counter term designed to ensure that eq. (152) is satisfied can now be added to eq. (175). We now want to show that upon expanding eq. (175) in powers of $x$ (noting that
$$ \not\!\!{C} = \frac{1}{x}\exp\left\lbrace \frac{2}{\sqrt{719}}\tan^{-1}\left[ \frac{10}{\sqrt{719}}\left(\frac{y}{x}- \frac{19}{10}\right)\right] \right\rbrace\;\;)\eqno(176)$$
does in fact give the $LL$ contribution to $V$ by comparing it with the results of ref. [25].

In ref. [25], these $LL$ contributions to $V$ are expressed in the form
$$V = \frac{\pi^6}{6} \phi^4 \left(y S_0 (yL) + \sum_{j=0}^\infty x^j L^{j-1} S_j (yL)\right).\eqno(177)$$
Direct expansion of eq. (175) in powers of $x$ to compare with the functions $S_0$ and $S_1$ occurring in eq. (177) that were computed in ref. [25] is awkward. We instead go back to the defining equation for $\overline{y}(t)$ in eq. (168) which implies that if
$$\overline{y}_0(t) = \overline{y}_0^{(0)} (t) + \overline{y}_0^{(1)} (t) x(\mu) + \ldots\eqno(178)$$
then
$$\frac{d\overline{y}_0^{(0)}}{dt} = \frac{5}{6} \overline{y}_0^{(0)}(t)\eqno(179)$$
and
$$\frac{d\overline{y}_0^{(1)}}{dt} = \frac{5}{6}\left(2\overline{y}_0^{(1)} \overline{y}_0^{(0)}\right) - 3\overline{y}_0^{(0)}.\eqno(180)$$
Solving eqs. (179-180) subject to the boundary conditions 
$\overline{y}_0^{(1)} = x\delta_{n0}$ leaves us with
$$\overline{y}_0(t) = \frac{y}{1 - \frac{5}{6} yt} + x\,\frac{
\frac{5}{4} y^2t - 3yt}{\left(1 - \frac{5}{6} yt\right)^2}+ O(x^2).\eqno(181)$$
Upon substitution of eqs. (171) and (181) into eq. (166) and taking $t$ to be given by eq. (174), we recover the first two contributions to eq. (177) found in ref. [25].  It is consequently reasonable to expect that eq. (175) gives the $LL$ approximation to $V$ to all orders in $x$. It should prove possible to obtain the full $LL$ approximation to $V$ in the standard model using this approach based on the method of characteristics. The implications of knowing $V$ exactly to $LL$ order on spontaneous symmetry breaking is clearly of interest.

\section{Discussion}

In this paper we have examined various features of the RG equation and how it can be used to determin portions of radiative effects in orders of perturbation theory beyond the order to which explicit calculations have been performed. Particular attention is paid to renormalization scheme dependency, the relationship between bare and renormalized quantities, the form of the general solution to the RG equation, and the way in which the method of characteristics can be used to extract $LL$ $NLL$ etc. effects from the RG equation.

\section{Acknowledgements}

Discussions with R.B. Mann and T.G. Steele were most helpful. R. and D. MacKenzie had useful suggestions, NSERC provided financial support. The hospitality of the Perimeter Institute where this work was completed is gratefully acknowledged.

\eject

\end{document}